\documentclass[runningheads]{llncs}
\usepackage{subcaption}
\usepackage{graphicx}
\usepackage{color}
\usepackage{soul,xcolor}
\definecolor{myblue}{RGB}{0,0,255}
\usepackage[hidelinks]{hyperref}
\hypersetup{
    colorlinks=true,
    urlcolor=blue,
}
\usepackage{booktabs}
\usepackage{multirow}
\usepackage{amsmath, amssymb, amsfonts}

\newcommand{\R}{\mathbb{R}}
\usepackage{tikz}
\newcommand{\tikzxmark}{%
\tikz[scale=0.18] {
    \draw[line width=0.7,line cap=round] (0,0) to [bend left=6] (1,1);
    \draw[line width=0.7,line cap=round] (0.2,0.95) to [bend right=3] (0.8,0.05);
}}
\begin{document}
\title{Let Me DeCode You: Decoder Conditioning with Tabular Data}
\author{Tomasz Szczepański\inst{1} \and
Michal K. Grzeszczyk \inst{1} \and
Szymon Płotka \inst{1,2,3} \and
Arleta Adamowicz \inst{4} \and
Piotr Fudalej \inst{4} \and
Przemysław Korzeniowski \inst{1} \and
Tomasz Trzciński \inst{5,6,7} \and
Arkadiusz Sitek \inst{8}
}
\authorrunning{T.Szczepański et al.}
%
\institute{Sano Centre for Computational Medicine, Cracow, Poland \\
\email{t.szczepanski@sanoscience.org} \and
Informatics Institute, University of Amsterdam, Amsterdam, The Netherlands \and
Amsterdam University Medical Center, Amsterdam, The Netherlands \and
Jagiellonian University Medical College, Cracow, Poland \and
Warsaw University of Technology, Warsaw, Poland \and
IDEAS NCBR, Warsaw, Poland \and
Tooploox, Wroclaw, Poland \and
Massachusetts General Hospital, Harvard Medical School, Boston, MA, USA \\
}
\maketitle              
\begin{abstract}
Training deep neural networks for 3D segmentation tasks can be challenging, often requiring efficient and effective strategies to improve model performance. In this study, we introduce a novel approach, DeCode, that utilizes label-derived features for model conditioning to support the decoder in the reconstruction process dynamically, aiming to enhance the efficiency of the training process. DeCode focuses on improving 3D segmentation performance through the incorporation of conditioning embedding with learned numerical representation of 3D-label shape features. Specifically, we develop an approach, where conditioning is applied during the training phase to guide the network toward robust segmentation. When labels are not available during inference, our model infers the necessary conditioning embedding directly from the input data, thanks to a feed-forward network learned during the training phase. This approach is tested using synthetic data and cone-beam computed tomography (CBCT) images of teeth. For CBCT, three datasets are used: one publicly available and two in-house. Our results show that DeCode significantly outperforms traditional, unconditioned models in terms of generalization to unseen data, achieving higher accuracy at a reduced computational cost. This work represents the first of its kind to explore conditioning strategies in 3D data segmentation, offering a novel and more efficient method for leveraging annotated data. Our code, pre-trained models are publicly available at \url{https://github.com/SanoScience/DeCode}.

\end{abstract}
\section{Introduction}

The annotation process in medical imaging is time-consuming, costly, and requires medical domain knowledge \cite{tajbakhsh2020embracing}. Furthermore, deep learning-based algorithms necessitate a large amount of annotated data for acceptable performance and generalization capabilities~\cite{litjens2017survey}. However, to enhance deep learning algorithms without relying solely on large-scale imaging data, the community explored the use of tabular features~\cite{huang2020fusion,xia2022automatic,plotka2023babynet++}.

In recent years, FiLM~\cite{perez2018film} has emerged, allowing adaptive influence on neural network intermediate features through feature-wise affine transformations based on conditioning information. Expanding on this concept, integrating tabular information has shown significant advantages for model performance~\cite{grzeszczyk2023tabattention,wolf2022daft,jacenkow2020inside}. An example of the beneficial integration is TabAttention mechanism~\cite{grzeszczyk2023tabattention}. This approach  incorporates biometric measurements which improve fetal birth weight estimation on ultrasound video scans. Similarly, DAFT~\cite{wolf2022daft} is proposed to conditionally shift and scale feature maps based on conditioning. Integration of a patient's clinical information with a 3D MRI image shows improvement in time-to-dementia prediction, underscoring the richness of Electronic Health Record (EHR) data. However, both methods primarily address regression problems where tabular data exhibits measurable correlations with the target task, and imaging features contribute to reducing estimation error.

\begin{figure}[t!]
    \centering
    \includegraphics[width=\textwidth]{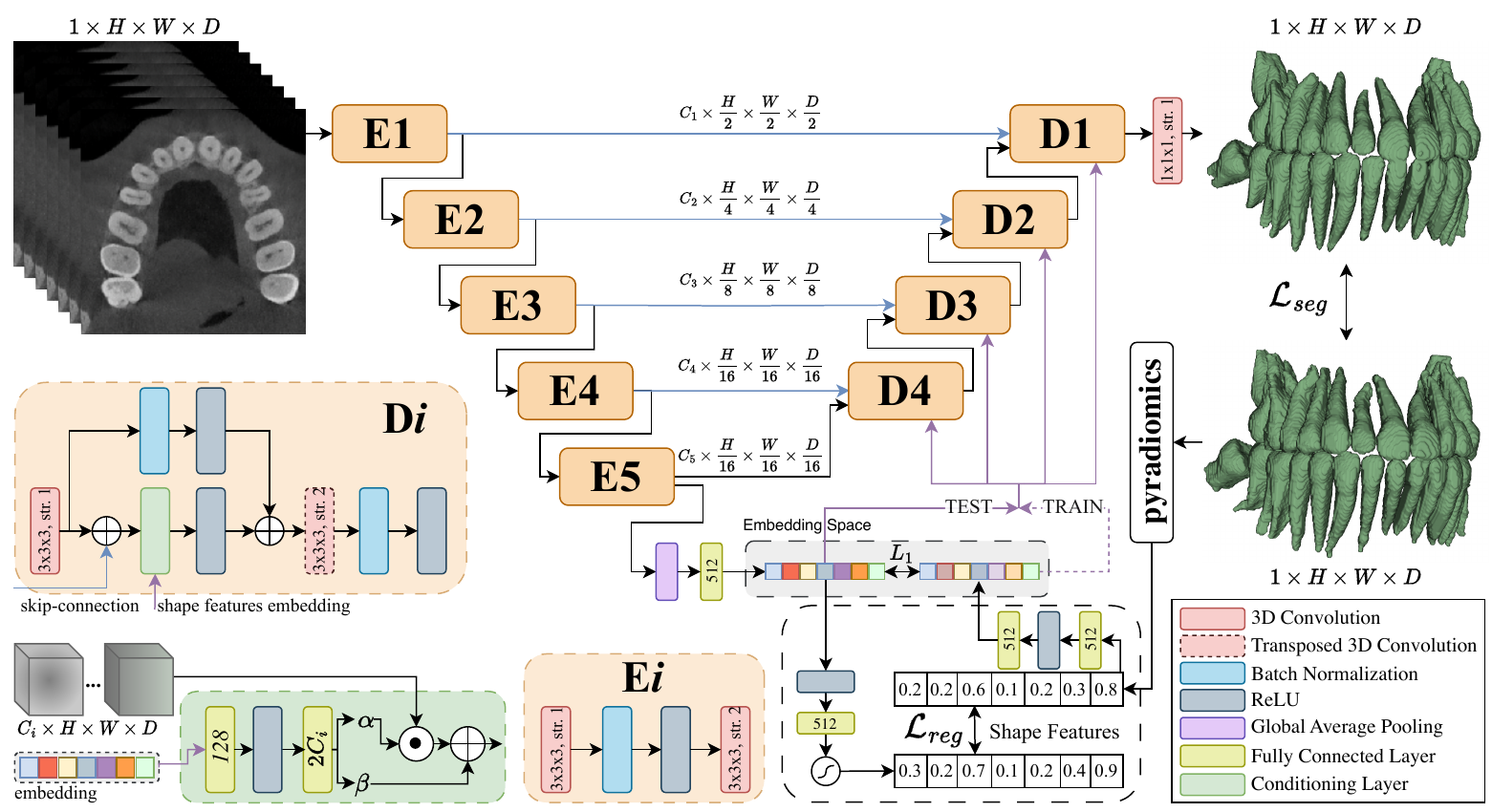}
    \caption{An overview of the proposed DeCode method for conditioning segmentation decoder with learned shape features embedding. During the inference when test labels are unavailable, we use the learned feature embedding optimized with $L_1$ loss in Eq.~\ref{eq:loss}. We perform conditioning after the skip connection from the encoder, allowing for a dynamic and selective decoding process. We also leverage a features regression as a helper task that boosts meaningful feature extraction. Skip connections and the flow of the shape features embedding are indicated with blue and purple arrows respectively. $Ei$ and $Di$ correspond to the encoder and decoder stages.}
    \label{fig:overview}
\end{figure}

Conversely, segmentation tasks often lack corresponding tabular data due to challenges such as fully anonymization process of medical data, the absence of a comprehensive data collection strategy, or clear relations between conditioning information and segmentation. The authors of conditioning layer INSIDE~\cite{jacenkow2020inside} propose to integrate non-imaging information into 2D segmentation network to improve performance. They utilize cardiac cycle phase and encoded 2D slice position as conditioning data. However, this prior-knowledge-based information relies on a simple two-state phase of the cardiac cycle, though correlated with the segmentation task, limits its information to a simple binary flag. This attempt sets out a line of research that we choose to pursue.

In this work, we explore conditioning on 3D data in a segmentation task when corresponding tabular data is unavailable. To our best knowledge, we are the first to investigate it. We introduce the DeCode, which performs conditioning based on shape feature embedding. For reproducibility, we calculate the label-based shape features using PyRadiomics~\cite{van2017computational}. We also demonstrate that the accompanying task of shape features regression benefits the model's segmentation performance and, most importantly, allows us to use feature embedding for conditioning during inference when test labels are unavailable. We evaluate our method on the novel synthetic DeCode 3D dataset, showing that shape features allow for conditioning synthetic tasks, thus demonstrating their usefulness. To demonstrate DeCode's applicability in a clinical setting, we use the 3D dental CBCT dataset~\cite{cui2022fully} to train the model and two external test sets to evaluate the generalization of the model. Accurate tooth delineation in dental CBCT images is essential for clinical diagnosis and treatment while preparing precise 3D labels is very time-consuming~\cite{polizzi2023tooth,zheng2024semi,li2022semantic,cui2021hierarchical,wang2023root}. Our conditioned architecture improves generalization to unseen data compared to the unconditioned one, which is trained on the same data while requiring no extra labeling work and only marginal additional training time. This work proposes a conditioning strategy for 3D data segmentation, offering a more efficient method for leveraging annotated data.

\section{Method}
In this section, we provide a detailed description of the network and the DeCode decoder with an emphasis on the application of conditioning information. Then, we describe the process of calculating shape features that are further used for the regression task to generate their embedding for inference-time conditioning.

The go-to standard for accomplishing medical imaging segmentation tasks is U-shaped architectures~\cite{minaee2021image,cui2022fully,ronneberger2015u,milletari2016v}. Here, we follow this approach and present lightweight architecture with an overview of our method in Fig.~\ref{fig:overview}. Let $X$ be a 3D CBCT scan $X\in \R^{1\times H \times W \times D}$ of height $H$, width $W$ and depth $D$, the U-shaped network generates multi-scale features at encoding stages $Ei$. Deeper encoder stages yield more abstract features up to the bottleneck within the deepest part of the architecture, containing compressed information from the input image. The decoding part $Di$ aims to reconstruct the segmentation map from features extracted by the encoder with additional skip connections at consecutive stages. Starting with high-level features of shape $14\times14\times10$ in the bottleneck through all decoder stages, we utilize learned feature embedding to condition the decoding process to improve the quality of the output mask.

\noindent
\textbf{Decoder Conditioning.} The first step within the decoding step is processing features from the previous stage with the convolutional layer. We add features from the encoder skip connections just before the conditioning layer to avoid leakage of low-level features without first conditioning them on shape feature embedding. The conditioning layer utilizes affine transformations to scale and shift feature maps. The transformation parameters are $\alpha_c$ and $\beta_c$, the products of hyper-network, which implement scale and offset, where $c$ is the number of feature map's channels (see Conditioning Layer in Fig.~\ref{fig:overview}). In contrast to FiLM conditioning, we parameterize the scale parameter to $(1 - \alpha_c)$ to facilitate the identity transform especially at the early stage of training, and to allow the scaling parameter to be regularized as a distance from zero. For $\alpha>0$, the scaling factor inverts a feature map, highlighting features that the ReLU activation would have otherwise suppressed \cite{rupprecht2018guide}. The conditioning operation takes a normalization role, replacing the batch norm between the convolutional layer and the activation function. In addition to the possibility of conditioning itself, this operation has an additional advantage compared to Batch Norm or Layer Norm: it does not depend on batch statistics~\cite{touvron2022resmlp}. The transformed features are summed via residual connection with the processed input to the decoder. The decoding step finishes with refining and up-sampling feature maps via transposed convolution, Batch Normalization, and a ReLU activation.

\begin{figure}[t!]
    \centering
    \includegraphics[width=1.0\textwidth]{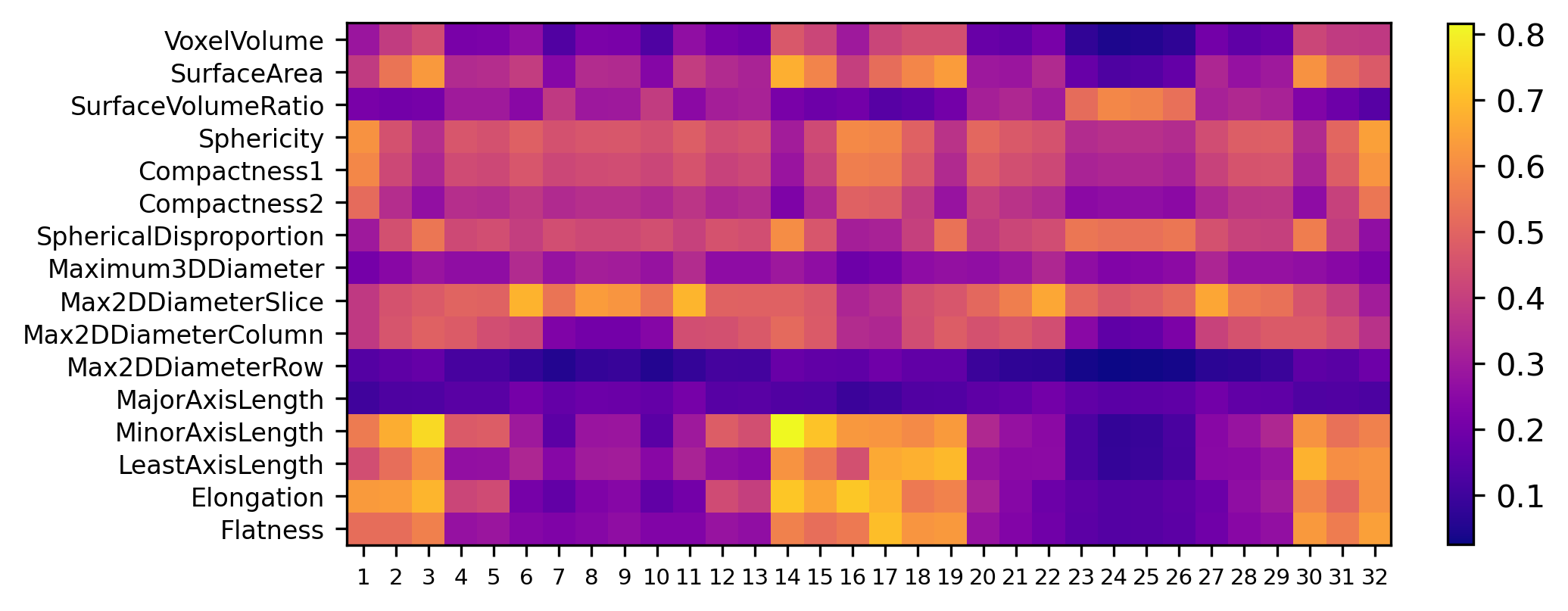}
    \caption{Normalized mean shape features calculated with PyRadiomics \cite{van2017computational} on CBCT Tooth dataset \cite{cui2022fully}. Each shape feature is calculated for every tooth separately revealing morphological differences between tooth types.}
    \label{fig:mean_shape_features}
\end{figure}

\noindent
\textbf{Shape Features.}  We utilize ground-truth masks to extract rich information and shape features. We consider, e.g., sphericity, volume, and elongation features, (see Fig.~\ref{fig:mean_shape_features}), which aim to decode more morphologically accurate masks. Before training, we extract shape features for every segmented object separately based on the ground truth mask (up to 32 objects, teeth, in a CBCT scan). This process yields a vector of length 512, forming tabular data that is used further to learn conditioning embedding. These features are utilized during training time to condition the decoding process. In real-world scenarios, the ground-truth masks are unavailable during inference. Therefore, we perform a shape-features regression task from the encoder's bottleneck latent space to replace the unavailable shape-features at the test stage with model-learned embedding (see Fig.~\ref{fig:overview}). For reproducibility and easy access, we calculate shape features with PyRadiomics~\cite{van2017computational}.

\noindent
\textbf{Loss function.} The multi-task loss function, which minimizes both segmentation, embedding distance, and regression tasks, is defined as follows:

\begin{equation}
\mathcal{L}=\mathcal{L}_{\text {Dice}}+ \Lambda_{1} \mathcal{L}_{\text{Focal}} + L_1 + \Lambda_{2}  \mathcal{L}_{\text{RMSE}} +  \eta ( \|\alpha\|_2^2 + \|\beta\|_2^2),
\label{eq:loss}
\end{equation}
\noindent
where $\Lambda_{1}$ = 0.5, and $\Lambda_{2}$ = 0.75. The coefficients are determined based on a trial-and-error optimization. $\mathcal{L}_{\text {Dice}}$ and $\mathcal{L}_{\text{Focal}}$ correspond to the segmentation task.  We optimize an $L_1$ distance to make encoder features embedding close to the representation of tabular shape features. During inference, we use this learned embedding to condition the decoder. We also add the helper task of shape features regression during training which we optimize based on Root Mean Square Error (RMSE). Finally, we add an $L_2$ penalty $\eta=0.00001$ to regularize the conditioning layer parameters $\alpha$ and $\beta$, due to the high capacity of the deep network following the conditioning layer, thus reducing the risk of overfitting.

\section{Experiments and Results}

In this section, we describe implementation details and introduce the synthetic dataset, 3DeCode, where we investigate the possibility of conditioning with shape features on 3D data. Moreover, we apply DeCode to the task of 3D segmentation, utilizing CBCT datasets. We highlight the significance of DeCode key components and evaluate its ability to generalize to unseen CBCT data in comparison to lightweight 3D UNet, which lacks decoder conditioning.

\noindent
\textbf{Implementation details.} We implement identical models for both synthetic and clinical datasets. We use a UNet network with 4-down and 4-up sampling stages, Batch Normalization, ReLU activations, and a Sigmoid layer for final classification. The conditioning layers are placed inside decoder stages as shown in Fig.~\ref{fig:overview}. We crop an ROI around the teeth based on labels with a size of 240$\times$240$\times$176 from the input CBCT scan. Then, we randomly crop a patch of size 224$\times$224$\times$160 and feed it to the network. We train the model using a batch size of 4, and the AdamW optimizer for 400 epochs. A learning rate is set to 0.001, and the weight decay is set to 0.0001. The intensity of the Hounsfield Unit is clipped to the range [0, 3500] and linearly scaled to [0, 1]. We employ geometric and intensity-related data augmentation such as random rotation, translation, or brightness and contrast adjustments throughout the training process. We implement our model in PyTorch 1.13.1 and MONAI 1.2.0 and train it on NVIDIA A100 80GB GPU with CUDA 11.6. We use PyRadiomics 3.1.0 to calculate binary mask shape features. In case of a missing tooth, we fill its shape features with a vector of zeros and finally normalize tabular data to a range of [0, 1]. We perform a paired t-test with p $<$ 0.05 to identify significant differences.

\setlength{\tabcolsep}{6pt}
\begin{table}[t!]\centering
\caption{Quantitative results on DeCode 3D dataset in the average Dice Similarity Coefficient (DSC) (\%). We explore the possibility of conditioning 3D solid with shape features in a segmentation task. For the tasks \textit{Size} and \textit{Shape}, we conditionally segment solids of small, medium, or large sizes and sphere, cube, or cylinder shapes, respectively. \textit{Mixed} task configurations combine the characteristics of \textit{Shape} and \textit{Size} tasks. More challenging \textit{Varying} combinations additionally address shape and size variability based on a uniform distribution, beyond binary combinations. We report the baseline as an unconditioned UNet.}
\label{tab:cond3decode}
\begin{tabular}{@{}lccc@{}}
\toprule
\multirow{2}{*}{Task} & \multicolumn{3}{c}{DSC (\%)}                                       \\ \cmidrule(l){2-4} 
                      & Baseline &  & Shape features conditioning \\\cmidrule(l){2-4} 
Size  & $49.18\pm32.26$    & & $98.23\pm3.92$                       \\
Shape & $53.48\pm22.09$    & & $99.33\pm0.85$                       \\
Mixed & $17.84\pm25.68$    & &  $97.96\pm5.25$                      \\
Varying Size &  $32.97\pm32.23$    & &    $97.48\pm5.69$                    \\ 
Varying Mixed &   $12.43\pm28.45$   & &     $94.74\pm12.94$                   \\ \bottomrule
\end{tabular}
\end{table}

\noindent
\textbf{3DeCode dataset.} We present a novel dataset inspired by CLEVR-Seg \cite{jacenkow2020inside}, extending it to 3D and generating segmentation masks based on conditioning scenario tasks. We design tasks that require conditioning based on Shape,  Size, or Shapes of different Sizes (referred to as Mixed). To utilize the rich information stored as non-binary shape features, we also enrich the dataset with solids of varying shapes and sizes. Namely, we generate two additional tasks that introduce non-discrete variability in Size or Shape to the solids, based on a uniform distribution, e.g., to generate the varying-size solid class 'small sphere' we vary its radius by $\pm$ 20\%. While this approach does not reflect the full spectrum of information that shape features can store, it allows us to assess the feasibility of conditioning on 3D data in a segmentation task. The Varying Mixed task consists of shapes varying in size and shape, where, e.g., the base spherical shape can result in an ellipsoid and a cube in a cuboid. The generated solids are binary, as complex image feature extraction is not a concern. Tasks to be solved accurately require the use of the conditioning information by the network. Otherwise, accuracy is reduced to a random guess based solely on the image. The positions of the solids are drawn randomly, whereby they may overlap. We generate 300 labeled conditions for tasks of Size (small, medium, or large) or Shape (sphere, cube, cylinder), and 900 for the Mixed tasks. Data consists of condition-based 3D images with up to 18 objects in volume space of the same size as the patch size used by our model. We generate every possible conditioning combination per image to prevent the model from memorizing image-condition pairs. For evaluation, we split the dataset into training, validation, and testing subsets with a 60:20:20 ratio. 3DeCode samples can be found in the supplementary material - Sec.\ref{sec:supplement}.

\noindent
\textbf{CBCT dataset.} To train our model, we use 98 publicly available 3D dental CBCT scans \cite{cui2022fully}. We evaluate the segmentation performance on an external test set, comprising 20 CBCT scans obtained from a retrospective study (IRB OKW-623/2022) conducted at two medical centers: Center A (11 scans) and Center B (9 scans). The ground truth annotations for the test set were performed by an orthodontist with 5 years of experience, who was verified by another orthodontist with 25 years of clinical practice. We resample all scans to 0.4 $\times$ 0.4 $\times$ 0.4 mm$^3$ isotropic resolution.

\noindent
\textbf{Results.} We present results on 3DeCode dataset in Table \ref{tab:cond3decode}. According to our dataset-building principle, the baseline UNet cannot segment the image without conditioning. The Mixed task DSC is 17\% which is close to a random sample 1 out of 9. The model with decoder conditioning can correctly perform the conditional 3D segmentation task, approaching perfect accuracy for the Shape task with a DSC of 99.23\%, and 94.74\%, respectively for the most challenging Varying Mixed task. Our model struggles only when overlapping solids, due to random placement solids, are present, which is unrelated to conditioning. The results of the experiment, demonstrate that it is possible to condition in 3D using shape features embedding, which allows us to move on to examine the impact of conditioning on clinical data segmentation. 

\setlength{\tabcolsep}{4pt}
\begin{table}[t!]\centering
\caption{Quantitative results on 3D CBCT datasets: external (Center A and Center B) and validation split. We report DSC and standard deviation. Configuration (1) refers to an unconditioned network serving as a baseline. An upper bound of generalization is provided by configuration (7) conditioned with shape features calculated on test-set masks. The proposed configuration DeCode (8) utilizes during test time learned feature embedding. We conduct a paired t-test to establish statistical significance between the baseline and configuration (7) and (8), denoted by (*) for $p <0.05$. CL stands for the Conditioning Layer, Reg the Regression task, CR Conditioning Information Representation, and T Test-time conditioning, Rand for Random Features, CSF for test-set Calculated Shape Features, and LESF for Learned Embedding of Shape Features.}
\label{tab:clinical_results}
\begin{tabular}{lcccclccccc}
\toprule
   & \multicolumn{4}{c}{Configuration}         &  & \multicolumn{5}{c}{DSC (\%)} \\ \cline{2-5} \cline{7-11} 
   & CL     & Reg    & CR & T   &  & Center A &  & Center B    &  & VAL \\ \hline
1. & -      & \tikzxmark           & -  & -  &  & $89.67\pm2.34$          &  & $94.55\pm1.16$           &  & $\mathbf{95.89\pm0.84}$ \\ 
2. & -      & \checkmark  & -   & - &  & $91.94\pm1.56$          &  & $95.41\pm1.01$           &  & $95.67\pm0.88$    \\ 
3. & FiLM   & \tikzxmark           & Rand & \tikzxmark  &  & $89.75\pm1.94$          &  & $93.72\pm1.33$           &  & $95.86\pm0.95$    \\ \midrule
4. & FiLM   & \tikzxmark           & CSF & \tikzxmark &  & $92.11\pm1.79$          &  & $95.45\pm1.12$  &  & $95.59\pm0.88$    \\
5. & INSIDE & \tikzxmark           & CSF & \tikzxmark &  & $91.16\pm3.33$          &  & $95.12\pm0.99$           &  & $95.61\pm0.73$    \\
6. & DAFT   & \tikzxmark           & CSF & \tikzxmark &  & $90.61\pm9.22$          &  & $95.14\pm1.05$           &  & $95.54\pm0.91$    \\
7. & FiLM   & \checkmark  & CSF & \tikzxmark &  & $\mathbf{93.12\pm1.07^*}$ &  & $\mathbf{95.52\pm0.92^*}$           &  & $95.60\pm0.93$    \\ \midrule
8. & FiLM & \checkmark  & LESF & \checkmark &  &   $ 92.74\pm1.34^*$                      &  &         $ 95.12\pm0.91^*$                 &  &     $95.81\pm0.86$            \\ \bottomrule
\end{tabular}
\end{table}

We compare the unconditioned UNet network's (1) results on the CBCT dataset, which serves as the baseline method, with the proposed DeCode (8) -- with numbers in brackets corresponding to configurations presented in Table~\ref{tab:clinical_results}. To find the optimal configuration, we explore the impact of an auxiliary shape feature regression task (2), different conditioning layers (CL) (4-6), and conditioning information representation (CR) (3-8). Firstly, we add a shape feature regression task (2) to the baseline method that improves generalization on both external sets, proving the usefulness of the shape features. Secondly, we evaluate conditioning layer types with calculated shape features (CSF), which, for this experiments, we also use during the test. We get the best results with the FiLM layer, so we use it for further experiments. We examine the edge case of conditioning on random tabular data generated from a standard normal distribution and observe a significant performance decline. The result for Center A is better than the unconditioned model, suggesting that the conditioning layer increases the model's capacity, posing a threat of overfitting. However, it may also suggest that residual connections in the decoder make the model robust to the possible negative impact of conditioning. A final experiment (7) based on the CSF leverages the FiLM layer and regression task. It sets an upper bound for generalization improvement. To adapt the method to test time, unlike configuration (7), we use learned embedding of shape features (LESF), which is our proposed configuration (8). Although the proposed DeCode does not improve the result on the validation set, it statistically significantly improves the generalization to new unknown data. Finally, we compare out method with unconditioned U-shaped networks (see Table \ref{tab:computation}). We choose architectures with a broad range of parameter numbers, provided they allow training with a large 3D patch under GPU memory constraints. Our solution achieves the second-best generalization, giving way only to the VNet method, which is, however, 10$\times$ more computationally intensive and requires 4$\times$ longer training.

\setlength{\tabcolsep}{1.5pt}
\begin{table}[t!]\centering
\caption{Performance comparison on CBCT test sets between baseline unconditioned U-shaped networks and the DeCode method. P denotes the number of parameters, I inference time, and T training time.}
\label{tab:computation}
\begin{tabular}{@{}llcccccccccc@{}}
\toprule
\multicolumn{1}{l}{\multirow{2}{*}{Network}} &  & \multirow{2}{*}{P (M)} & \multirow{2}{*}{GFLOPs} & \multirow{2}{*}{I (ms)} & \multirow{2}{*}{T (h)} & \multicolumn{1}{l}{} & \multicolumn{5}{c}{DSC (\%)} \\
\multicolumn{1}{c}{} &  &  &  &  &  & \multicolumn{1}{l}{} & Center A &  & Center B &  & Avg. \\ \midrule
UNext \cite{valanarasu2022unext} &  & 4 & 46 & 21 & 1.5 &  & $88.98\pm3.32$ &  & $93.37\pm1.80$ &  & $90.96\pm2.64$ \\
UNet \cite{ronneberger2015u} &  & 25 & 1880 & 117 & 8.5 &  & $92.03\pm1.45$ &  & $94.41\pm0.98$ &  & $93.10\pm1.24$ \\
ResUNet34 \cite{he2016deep} &  & 70 & 2610 & 101 & 11 &  & $92.28\pm1.32$ &  & $95.56\pm0.99$ &  & $93.71\pm1.17$ \\
Att-UNet \cite{oktay2018attention} &  & 6 & 380 & 127 & 5 &  & $92.66\pm1.51$ &  & $95.22\pm1.06$ &  & $93.81\pm1.31$ \\
VNet \cite{milletari2016v} &  & 46 & 2770 & 175 & 13 &  & $93.07\pm0.93$ &  & $95.42\pm1.02$ &  & $94.13\pm0.97$ \\ \midrule
DeCode &  & 4 & 204 & 41 & 3 &  & $92.74\pm1.34$ &  & $95.12\pm0.91$ &  & $93.81\pm1.15$ \\ \bottomrule
\end{tabular}
\end{table}

\section{Conclusions}
This paper investigates the possibility of conditioning the decoder in the 3D segmentation task on the tabular data. Compared to unconditioned training, DeCode performs better on unseen data, requiring no extra labeling work and marginal additional training time. We evaluated our method on two external CBCT datasets, proving its enhanced generalizability. Obtained results encourage further research in this field, allowing more efficient use of annotated data.

There are limitations to our method. Firstly, we train our method on a relatively small dataset where selecting hyperparameters is complex, and their small changes may lead to a loss of stability in embedding learning, including their collapse. We expect better stability and further segmentation improvement with the increased dataset. Secondly, the radiomics features provide information limited to shape without considering objects' positions and relations between them. In the future, we plan to conduct the conditioning on features extracted automatically from labels, enabling the end-to-end training of representations for improved clinical image segmentation.

\begin{credits}
\subsubsection{\ackname} This work is supported by the EU's Horizon 2020 programme (grant no. 857533, Sano) and the Foundation for Polish Science's International Research Agendas programme, co-financed by the EU under the European Regional Development Fund and the Polish Ministry of Science and Higher Education (contract no. MEiN/2023/DIR/3796). This research was funded in whole or in part by National Science Centre, Poland 2023/49/N/ST6/01841. For the purpose of Open Access, the author has applied a CC-BY public copyright licence to any Author Accepted Manuscript (AAM) version arising from this submission.

\subsubsection{\discintname}
The authors have no competing interests to declare.
\end{credits}

%
%

\bibliographystyle{splncs04}
\bibliography{DeCode}
\clearpage
\section*{Supplementary material}
\label{sec:supplement}

\subsection*{3DeCode synthetic dataset}

\begin{figure}
    \centering
    \includegraphics[width=\textwidth]{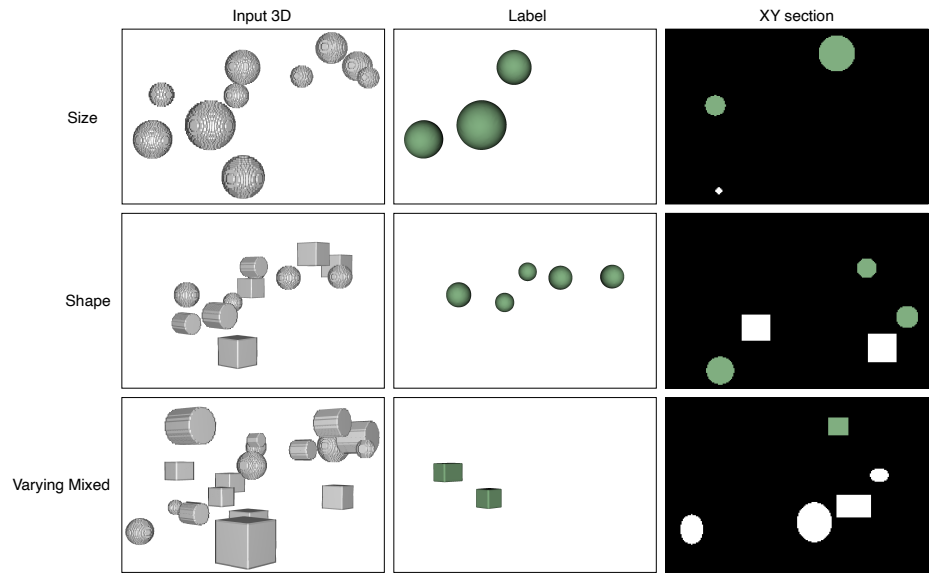}
    \caption{The 3DeCode data samples. The first column presents a 3D image, the basis for various configurations corresponding to the conditioning task, given along rows. Exemplary labels can be found in the central column. In the last column, we present one of the cross-sections. The dataset can be generated using the provided source code and attached configuration files with seeds.}
\end{figure}

\clearpage
\subsection*{Radiomics shape features}
\begin{figure}[h]
    \centering
    \caption{Normalized mean shape features calculated with PyRadiomics on the proprietary test datasets. Each shape feature is calculated for every tooth separately revealing morphological differences between tooth types. A small difference in mean values between the datasets shape features can be found.}
\begin{subfigure}{\textwidth}
    \includegraphics[width=\textwidth]{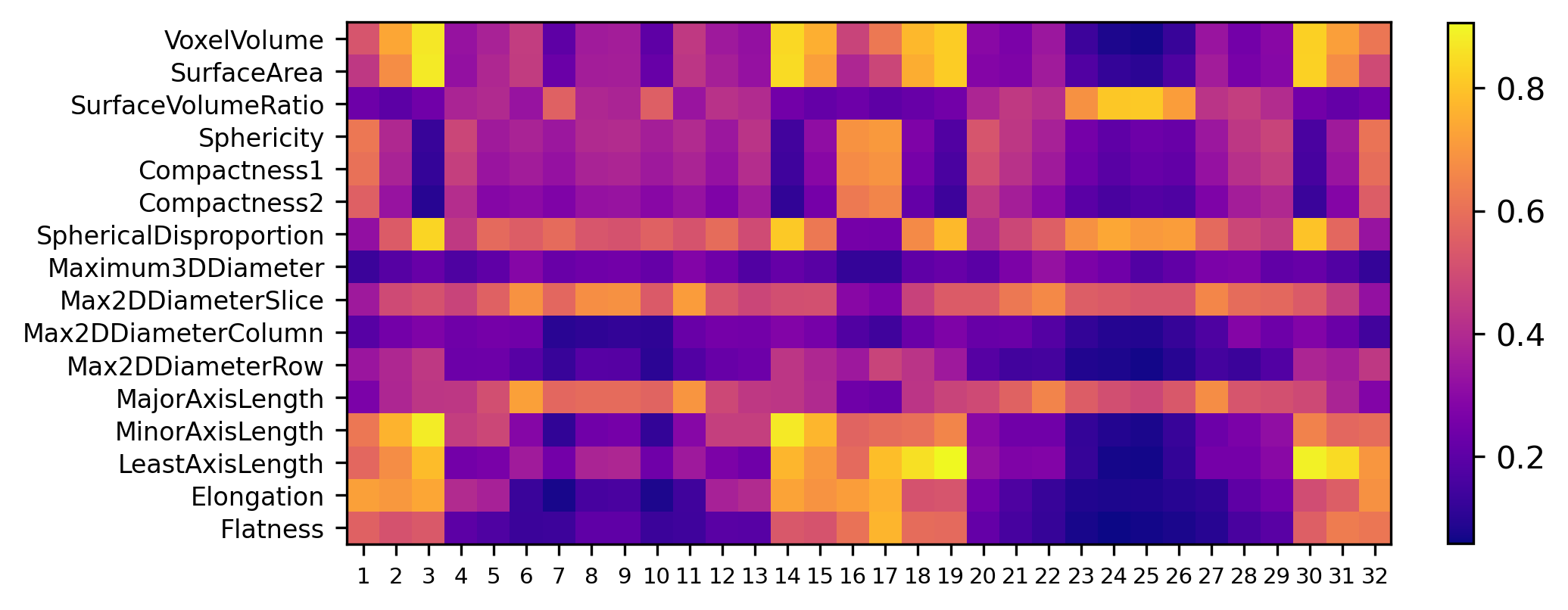}
    \caption{Dataset Center A.}
    \label{fig:first}
\end{subfigure}
\hfill
\begin{subfigure}{\textwidth}
    \includegraphics[width=\textwidth]{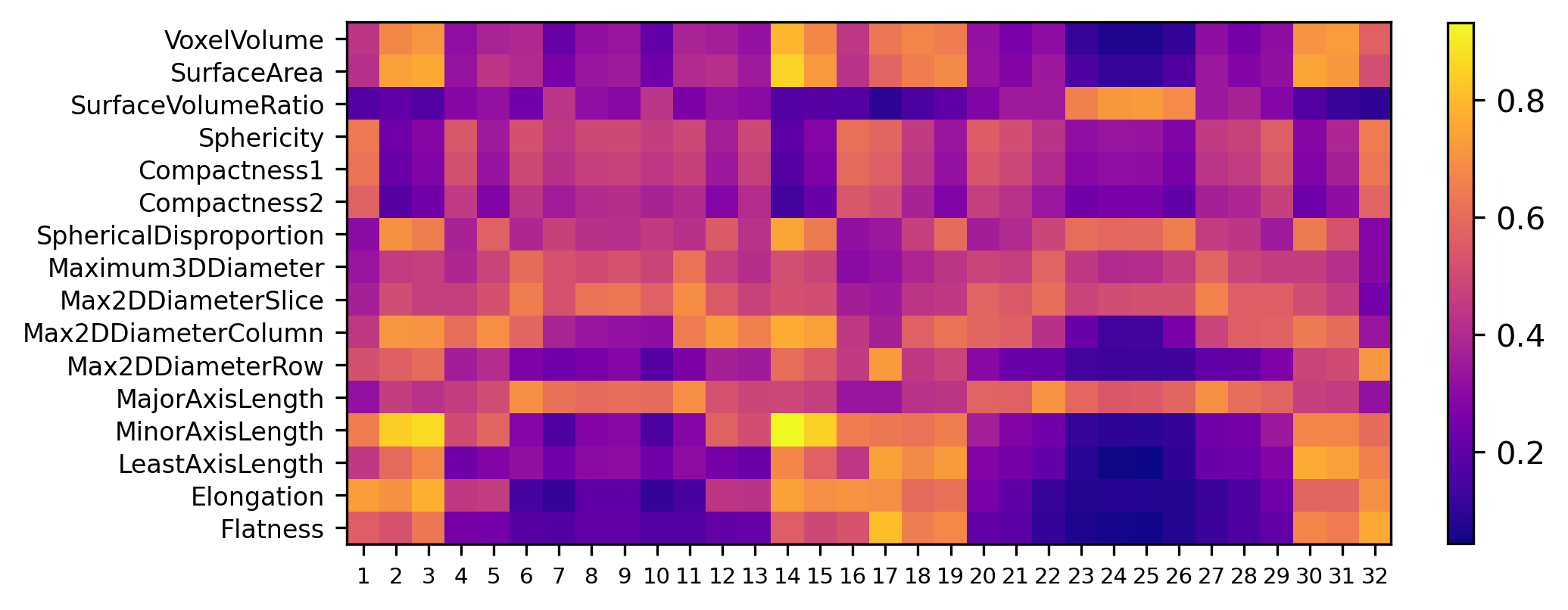}
    \caption{Dataset Center B.}
    \label{fig:third}
\end{subfigure}
\end{figure}

\end{document}